\documentclass[letterpaper]{IEEEtran}
\usepackage{cite}

\ifCLASSINFOpdf
  \usepackage[pdftex]{graphicx}
\else
\fi
\usepackage{amsmath}
\usepackage{amssymb}
\emergencystretch=\maxdimen
\usepackage{booktabs}
\usepackage{multirow}
\newcommand{\tabincell}[2]{\begin{tabular}{@{}#1@{}}#2\end{tabular}}

\usepackage{array}

 \usepackage[caption=false,font=footnotesize]{subfig}
\begin{document}
%
\title{A Practical Solution for SAR Despeckling \\ With Adversarial Learning Generated Speckled-to-Speckled Images\vspace{0.2cm}}

%
%
%

\author{Ye~Yuan, \IEEEmembership{Student Member, IEEE},
       Jian~Guan, \IEEEmembership{Member, IEEE},
        Pengming~Feng, \IEEEmembership{Member, IEEE},
		and~Yanxia~Wu
\thanks{}
\thanks{}
\thanks{}\thanks{This work was supported in part by the Fundamental Research Funds for the Central Universities under Grant 3072020CFT0602, in part by the Open Research Fund of State Key Laboratory of Space Ground Integrated Information Technology under Grant\_2018\_SGIIT\_KFJJ\_AI\_01, and in part by the National Natural Science Foundation of China under Grant 61806018. (\textit{Corresponding author: Jian~Guan.})}
\thanks{Ye Yuan, Jian Guan, and Yanxia Wu are with the College of Computer Science and Technology, Harbin Engineering University, Harbin 150001, China (e-mails: yuanye@hrbeu.edu.cn, j.guan@hrbeu.edu.cn, wuyanxia@hrbeu.edu.cn).}
\thanks{Pengming Feng is with the State Key Laboratory of Space-Ground Inte-grated Information Technology, China Academy of Space Technology, Beijing 100086, China (e-mail: p.feng.cn@outlook.com).}
\thanks{}\vspace{-0.3cm}}

\maketitle

\begin{abstract}
  \label{abs}
  In this letter, we aim to address a synthetic aperture radar (SAR) despeckling problem with the necessity of neither clean (speckle-free) SAR images nor independent speckled image pairs from the same scene, and a practical solution for SAR despeckling (PSD) is proposed. First, an adversarial learning framework is designed to generate speckled-to-speckled (S2S) image pairs from the same scene in the situation where only single speckled SAR images are available. Then, the S2S SAR image pairs are employed to train a modified despeckling Nested-UNet model using the Noise2Noise (N2N) strategy. Moreover, an iterative version of the PSD method (PSDi) is also presented. Experiments are conducted on both synthetic speckled and real SAR data to demonstrate the superiority of the proposed methods compared with several state-of-the-art methods. The results show that our methods can reach a good tradeoff between feature preservation and speckle suppression.
\end{abstract}


%
\IEEEpeerreviewmaketitle

\section{Introduction}
%
%
%
%
\IEEEPARstart{S}{ynthetic} aperture radar (SAR) images are usually corrupted with speckle noise, which leads to the degrada-tion of image quality and affects the performance in various applications of remote sensing~\cite{lukin2018despeckling}, e.g., classification~\cite{mahdianpari2017effect} and change detection~\cite{wang2019can}. Several methods have been proposed to mitigate the speckle in SAR images, including filter-ing methods~\cite{lee_digital_1980}, \cite{kuan1985adaptive}, wavelet shrinkage~\cite{chang2000adaptive,li2012bayesian}, variational models \cite{aubert2008variational, shi2008nonlinear, chen2014higher}, and SAR block-matching 3-D algorithm (SAR-BM3D)~\cite{parrilli_nonlocal_2012}. However, these methods sometimes fail to preserve sharp features such as edges or contain block artifacts in the despeckled images~\cite{argenti2013tutorial}.

Recently, convolutional neural network (CNN)-based supervised learning has been employed for SAR despeckling and achieved remarkable despeckling performance~\cite{chierchia_sar_2017,wang_sar_2017,zhang_learning_2018, ma2019cnn}, which can reduce speckle noise by learning relationships between speckled and clean ground truth images with CNN models.
However, there are few clean SAR images in practi-cal applications. Hence, speckle-free optical images with the single channel are usually employed as the clean ground truth for SAR image despeckling, and synthetic speckled images can be obtained by adding speckle noise to original images. Then, such speckled-to-clean image pairs can be used as training data in CNN despeckling models. However, because of the differences in imaging mechanism and image features of SAR and optical images, i.e., gray-level distribution and spatial correlation~\cite{di2013benchmarking}, it is not the optimal solution to achieve SAR image despeckling by directly employing the aforementioned CNN models, which are trained on optical images using supervised learning.

More recently, the Noise2Noise (N2N)~\cite{Lehtinen2018Noise2NoiseLI} strategy has shown its ability for image denoising without using any clean ground-truth images. By employing the N2N strategy, the CNN model can still achieve high denoising perfor-mance (e.g., zero-mean distribution noise) with mean square error (MSE) loss, as long as two independent noisy images from the same underlying clean image are available. This provides the possibility for SAR image despeckling without using clean ground-truth images. However, to acquire two independent speckled SAR images from the same scene is quite difficult. Hence, it is still impossible to solve the SAR image despeckling problem by merely employing the N2N strategy, as only single speckled SAR images are available.

In this letter, we introduce a practical despeckling solution for SAR image in the situation of only single speckled images are available, namely, a practical solution for SAR despeckling (PSD). Our contributions are summarized as follows.
\vspace{0.5cm}
\begin{figure*}[tbp]
	\centering
	\includegraphics[width=0.60\textwidth]{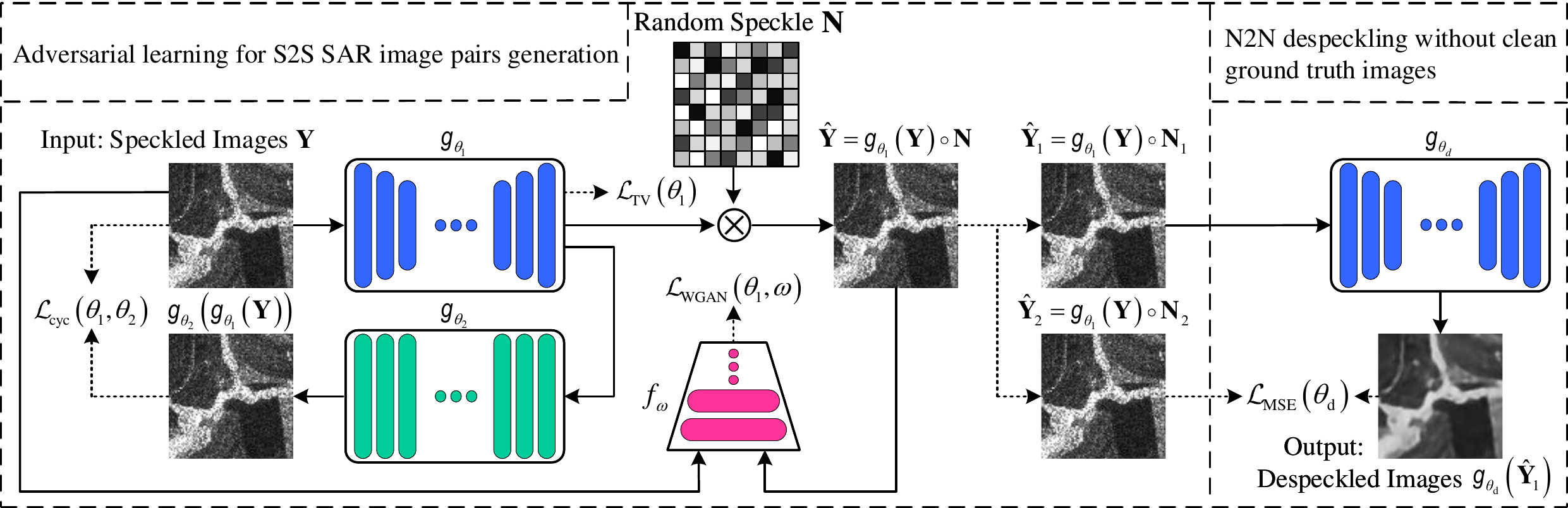}
	\caption{Overall flowchart of the proposed PSD method. (Left) Adversarial learning for S2S SAR image pairs generation: two generators ${g_{{\theta _1}}}$ and ${g_{{\theta _2}}}$ and a discriminator $f_\omega$ are trained by using the adversarial loss ${{\mathcal L}_{{{\text{WGAN}}}}}$, the backward cycle consistency loss $\mathcal{L}_{\text{cyc}}$, and the TV loss $\mathcal{L}_{{\text{TV}}}$.
		(Right) N2N despeckling without clean ground-truth images: the despeckling model ${g_{{\theta _d}}}$ is trained by using the MSE loss ${{\cal L}_{{\rm{MSE}}}}$. $ \otimes $ denotes adding multiplicative speckle noise according to \eqref{eq:1} and \eqref{eq:1-1}.}
	\label{generate_model}
	\vspace{-0.2cm}
\end{figure*}

\begin{enumerate}[]
  \item To generate speckled-to-speckled (S2S) SAR image pairs, an adversarial learning framework that consists of two generators and a discriminator is presented, which are trained by an alternating optimization strategy.
  \item  By using the obtained S2S SAR image pairs, an advanced Nested-UNet model~\cite{zhou2018unetjia} is trained to achieve despeckling with the N2N strategy.
  In addition, an iterative version of the PSD method (PSDi) is proposed, which can further improve the despeckling performance.
  \item Visual and quantitative experiments conducted on synthetic speckled and real SAR data show that the proposed methods notably suppress speckle noise with better preserving features, which outperform several state-of-the-art despeckling methods.
\end{enumerate}

\section{Methodology}

In this section, we present implementation details of the proposed PSD method as shown in Fig.~\ref{generate_model}, which consists of two parts: 1) adversarial learning framework for S2S SAR image pairs generation and its loss functions for optimization and 2) N2N despeckling strategy without using any clean ground-truth images and the despeckling network architecture of the modified Nested-UNet.

\subsection{SAR Speckle Noise Model }
\label{speckle_model}
Let ${\textbf{Y} \in \mathbb{R}^{W \times H}} $ be the observed SAR image with the size of ${W \times H}$, ${\textbf{X} \in \mathbb{R}^{W \times H}} $ be the underlying (despeckled) clean image, and ${\textbf{N} \in \mathbb{R}^{W \times H}} $ be the speckle noise. Then $\textbf{Y}$ can be obtained by the following multiplicative model~\cite{wang_sar_2017}:
\begin{equation}
	{{\textbf{Y}}={\textbf{X}} \circ {\textbf{N}}}
	\label{eq:1}
\end{equation}%
where $\circ$ denotes the Hadamard product (i.e., entrywise prod-uct) of two matrices and ${\textbf{N}}$ is assumed to follow a gamma distribution with unit mean. The probability density function of ${\textbf{N}}$ can be written as~\cite{wang_sar_2017}:
\begin{equation}
p\left( \textbf{N} \right) = {{{L^L}{\textbf{N}^{L - 1}}{e^{ - L\textbf{N}}}} \over {\Gamma \left( L \right)}}.
\label{eq:1-1}
\end{equation}%
Here, $\Gamma \left( \cdot \right)$ denotes the gamma function and $L$ is the number of looks in SAR imaging process, where a smaller $L$ indicates stronger speckle. In this work, the approach of adding speckle noise is following the scheme in~\eqref{eq:1} and \eqref{eq:1-1}. 
The aim of despeckling is to estimate $\textbf{X}$ from the observed image $\textbf{Y}$. Based on the established speckle noise model in~\eqref{eq:1} and \eqref{eq:1-1}, our proposed approach in Section~\ref{Generating} can be used to generate S2S SAR image pairs from the same scenario.

\subsection{Adversarial Learning for S2S SAR Image \\Pairs Generation}
\label{Generating}

To generate S2S SAR image pairs, we introduce an adver-sarial learning framework, which consists of two generators ${{g}_{{\theta_1}}}$ and ${{g}_{{\theta_2}}}$, and a discriminator $f_\omega$, as shown in Fig.~\ref{generate_model} (left). $\theta_1$, $\theta_2$ and $\omega$ denote the parameters (i.e., weights and biases) of ${g_{{\theta _1}}}$, ${g_{{\theta _2}}}$ and $f_\omega$, respectively. ${g_{{\theta _1}}}$ is used to generate the ``fake" speckled SAR images $\hat {\textbf{Y}}$, expressed as follows:
\begin{equation}
\hat {\textbf{Y}} =  {{g_{{\theta _1}}}({{\textbf{Y}}}) \circ  {{\textbf{N}}}}.
\end{equation}
The network architecture of ${g_{{\theta _1}}}$ is the same as the despeckling model ${g_{{{{\theta }}_d}}}$, which will be described in Section~\ref{Iterative}. Also, the discriminator ${f_\omega  }$ determines the distance between the distribution of the ``fake" speckled SAR images and the distribution of the input speckled images, and its network architecture is designed as suggested in \cite{cha2019gan2gan}. ${g_{{\theta _1}}}$ and $f_\omega$ are trained by the adversarial loss of Wasserstein generative adversarial networks (WGANs)~\cite{arjovsky2017wasserstein} ${{\mathcal L}_{{{\text{WGAN}}}}}$, which can be formulated as
\begin{equation}
\label{lossWGAN}
{{\mathcal L}_{{{\text{WGAN}}}}} \left( {{\theta _1},{\omega} } \right)  = {{\mathbb{E}}}\left[ {{f_\omega }({{\textbf{Y}}})} \right] - {{\mathbb{E}}}\left[ {{f_\omega  }\left( {{g_{{\theta _1}}}({{\textbf{Y}}}) \circ  {{\textbf{N}}}} \right)} \right],
\end{equation}
where $\mathbb{E}$ denotes the expectation operator.

To make $g_{{\theta}_{1}}$ change speckle distribution while preserving the basic edge features of the original SAR image, $g_{{\theta}_{2}}$ is used to bring the output of $g_{{\theta}_{1}}$ back to the original SAR image, namely, $\textbf{Y} \to {g_{{\theta _1}}}\left( \textbf{Y} \right) \to {g_{{\theta _2}}}\left( {{g_{{\theta _1}}}\left( \textbf{Y} \right)} \right) \approx \textbf{Y}$. We design the network architecture of $g_{{\theta}_{2}}$ following the denoising CNN (DnCNN)~\cite{zhang2017beyond}. ${g_{{\theta _1}}}$ and ${g_{{\theta _2}}}$ are trained by the backward cycle consistency loss $\mathcal{L}_{\text{cyc}}$ following \cite{cha2019gan2gan} and \cite{zhu2017unpaired}, denoted as
\begin{equation} \label{losscyc}
\mathcal{L}_{\text{cyc}}\left({\theta}_{1}, {\theta}_{2}\right) = \mathbb{E}\left[\left\|\textbf{Y}-g_{{\theta}_{2}}\left(g_{{\theta}_{1}}(\textbf{Y})\right)\right\|_{1}\right].
\end{equation}

In addition, to smooth the output of ${g_{{\theta _1}}}$, the generator ${g_{{\theta _1}}}$ is also trained by the total variation (TV) loss $\mathcal{L}_{{\text{TV}}}$, which is defined as%
\begin{equation}
\label{lossTV}
\begin{split}
	{{\cal L}_{{\text{TV}}}}( {{\theta _1}} )=\!\!
	{\sum \limits_{w = 1}^{W-1}} {\sum \limits_{h = 1}^{H-1}} 
	(&{{{| {{g_{{\theta _1}}}{{({\textbf{Y}})}_{w + 1,h}} - {g_{{\theta _1}}}{{({\textbf{Y}})}_{w,h}}} |}^2}}\\
	+&{{{| {{g_{{\theta _1}}}{{({\textbf{Y}})}_{w,h + 1}} - {g_{{\theta _1}}}{{({\textbf{Y}})}_{w,h}}} |}^2}})^{\!1/2}
\end{split}
\end{equation}
where ${{{{g_{{\theta _1}}}({\textbf{Y}})}}_{w,h}}$ is the pixel value in ${ {{g_{{\theta _1}}}({\textbf{Y}})} }$ and $\mathcal{L}_{{\text{TV}}}$ can reduce the difference of adjacent pixel values in the despeckled images ${ {{g_{{\theta _1}}}({\textbf{Y}})} }$.

With the defined loss functions, an alternating optimization strategy is applied to optimize ${g_{{\theta _1}}}$, ${g_{{\theta _2}}}$, and  $f_\omega$, which can be described as an adversarial min-max problem, expressed as
\begin{equation}
\label{minmax}
\min _{{\theta}_{1}, {\theta}_{2}} \max _{{\omega}}\left[ \mathcal{L}_{\text{WGAN}}\!\left({\theta}_{1}, {\omega}\right) \!+\!  \mathcal{L}_{\text{cyc}}\!\left({\theta}_{1}, {\theta}_{2}\right) \!+\! \alpha \mathcal{L}_{\text{TV}}\!\left({\theta}_{1}\right)\right]
\end{equation}
where $\alpha$ is the predefined weight for $\mathcal{L}_{{\text{TV}}}$. To prevent ${ {{g_{{\theta _1}}}({\textbf{Y}})} }$ from being oversmooth, $\alpha$ should be far less than 1~\cite{wang_sar_2017}.
After reaching a steady state through adversarial learning, i.e., until the discriminator $f_\omega$ cannot distinguish the ``fake" speckled SAR images ${{g_{{\theta _1}}}({{\textbf{Y}}}) \circ  {{\textbf{N}}}}$ from the input speckled images $\textbf{Y}$, then we can obtain the S2S SAR image pairs $( {\hat {\textbf{Y}}_1,\hat {\textbf{Y}}_2} ) $ from the same scene in the situation when only single speckled SAR images are available, which can be expressed as follows:
\begin{equation}
\label{s2s_image_pairs}
( {\hat {\textbf{Y}}_1,\hat {\textbf{Y}}_2} ) = \left( {{g_{{{{\theta }}_1}}}\left( {{{\textbf{Y}}}} \right) \circ  {{\textbf{N}}}_1,{g_{{{{\theta }}_1}}}\left( {{{\textbf{Y}}}} \right) \circ  {{\textbf{N}}}_2 }\right)
\end{equation}%
where ${{\textbf{N}}}_1$ and ${{\textbf{N}}}_2$ are two independent speckle matrices with the same number of looks $L$.
\subsection{N2N Despeckling Without Clean Ground-truth Images}
\label{Iterative}

\begin{figure}[t]
	\centering
	\includegraphics[width=6.4 cm]{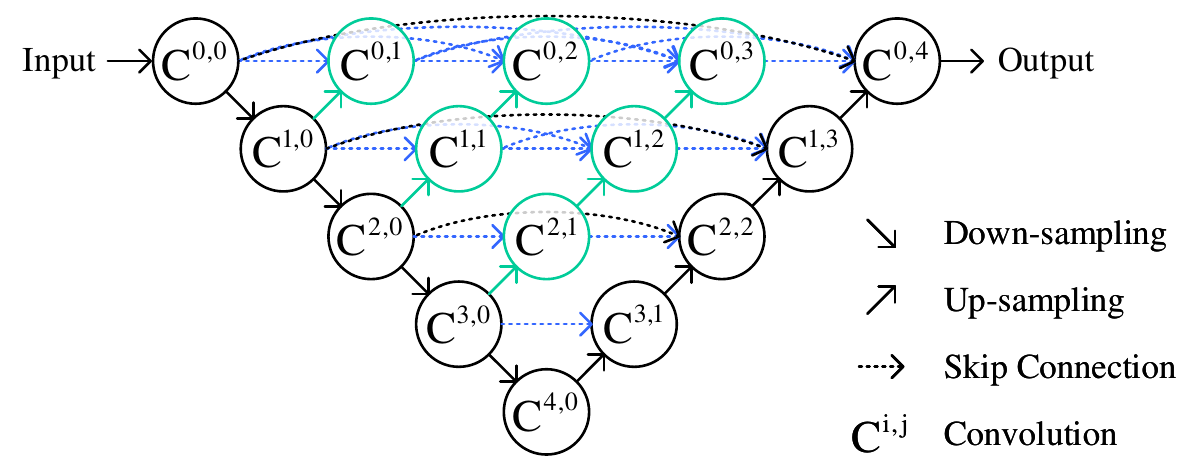}
	\caption{Network architecture of Nested-UNet.}
	\label{fig_N2N_Net}
\end{figure}
After obtaining the S2S SAR image pairs $( {\hat {\textbf{Y}}_1,\hat {\textbf{Y}}_2} )$, we can employ the N2N strategy~\cite{Lehtinen2018Noise2NoiseLI} to train the despeckling model ${g_{{{{\theta }}_d}}}$ without using any clean ground truth images.
Here, MSE loss is used to optimize ${g_{{{{\theta }}_d}}}$, which is formulated as 
\begin{equation}
{
	{\cal L}_{{\text{MSE}}}({{{\theta }}_d})={{|| {{\hat {\textbf{Y}}}_2 - g_{{\theta}_{d}}({\hat {\textbf{Y}}}_1)} ||}_2}
}
\end{equation}%
where $\theta_d$ denotes the parameters (i.e., weights and biases) of ${g_{{{{\theta }}_d}}}$. 
Due to the characteristic of speckle noise, as expressed in \eqref{eq:1} and \eqref{eq:1-1}, the expected value of speckled image is the same as that of underlying despeckled image, which is expressed as follows:
\begin{equation}
	{{\mathbb{E}}}[ {{\hat {\textbf{Y}}}} ]= {{\mathbb{E}}}\left[ {g_{{{{\theta }}_1}}}\left( {{{\textbf{Y}}}} \right)\circ  {{\textbf{N}}} \right]={{\mathbb{E}}}\left[{g_{{{{\theta }}_1}}}\left( {{{\textbf{Y}}}} \right) \right]\approx {{\mathbb{E}}}\left[{{{\textbf{X}}}}\right].
\end{equation}%
Hence, the minimum of ${\cal L}_{{\text{MSE}}}$ can be found at the expecta-tion of speckled images. This is particularly true when ${g_{{{{\theta }}_d}}}$ is trained on a large data set. Therefore, we can generate despeckled images by using ${g_{{{{\theta }}_d}}}$ without clean ground truth images.

A modified Nested-UNet~\cite{zhou2018unetjia} is adopted as the despeckling model ${g_{{\theta _{d}}}}$, as shown in Fig.~\ref{fig_N2N_Net}. Nested-UNet is chosen because of its advantage in improving the gradients flow throughout the network by its convolution layers on skip pathways and dense skip connections on skip pathways. The original Nested-UNet is polished to be more effective for SAR image despeckling by removing batch normalization layers of each convolutional block, which has been proven in other image processing tasks, e.g., image super-resolution~\cite{lim_enhanced_2017}.

Except for the above PSD method, we also propose an iterative version of PSD method, named PSDi.
For PSD method, the despeckling network is trained by using the generative S2S SAR image pairs $( {\hat {\textbf{Y}}_1, \hat {\textbf{Y}}_2} )$, and hence, their quality will affect the despeckling performance. The quality of the image pairs $( {\hat {\textbf{Y}}_1,\hat {\textbf{Y}}_2} )$ can be described as how close is the distribution of $( {\hat {\textbf{Y}}_1,\hat {\textbf{Y}}_2} ) $ to that of the real speckled SAR image $\textbf{Y}$. According to \eqref{eq:1} and \eqref{s2s_image_pairs}, we know $( {\hat {\textbf{Y}}_1,\hat {\textbf{Y}}_2} ) $ are generated by using ${ {{g_{{\theta _1}}}} }$, and their quality is decided by the despeckling ability of ${ {{g_{{\theta _1}}}} }$. However, the despeckling ability of ${g_{{\theta _1}}}$ is not decent, as there are no available ground-truth (clean or speckled) images, whereas ${g_{{{{\theta }}_d}}}$ can provide more powerful despeckling ability with ground-truth (speckled) images. Thus, to improve the quality of generative S2S SAR image pairs, we replace ${g_{{\theta _1}}}$ with ${g_{{{{\theta }}_d}}}$ in \eqref{s2s_image_pairs} and obtain the new S2S image pairs as follows:
\begin{equation}
\label{new_s2s_image_pairs}
( {\tilde {\textbf{Y}}_1,\tilde {\textbf{Y}}_2} ) = \left( {{g_{{{{\theta }}_d}}}\left( {{{\textbf{Y}}}} \right) \circ  {{\textbf{N}}}_1,{g_{{{{\theta }}_d}}}\left( {{{\textbf{Y}}}} \right) \circ  {{\textbf{N}}}_2 }\right).
\end{equation}
With the new S2S image pairs, a more effective despeckling model ${g_{{\theta _{di}}}}$ can be obtained by using N2N strategy again. Thereby, we can complete the training process of the despeck-ling model with only single speckled SAR images.

\section{Experiments and Results}
\vspace{0.2cm}
\begin{table}[tb]  
	\caption{{Performance Comparison in Terms of PSNR and SSIM on Synthetic Speckled Data}}
	\centering
  \scriptsize
  \begin{tabular}{ccccccc}
	\toprule
	\multirow{2}{*}{\tabincell{c}{The number \\of looks}}   & \multicolumn{3}{c}{PSNR} & \multicolumn{3}{c}{SSIM} \\
	\cmidrule(r){2-4} \cmidrule(r){5-7}
	 & 1      & 4      & 16     & 1      & 4      & 16     \\ 
	\midrule
	Speckled Input  & 11.08  & 15.17  & 20.73  & 0.0519 & 0.1261 & 0.3052 \\
	Chen2014  & 13.64  & 13.27  & -      & 0.4212 & 0.1651 & -      \\
	SAR-BM3D & 16.99  & 19.88  & 22.24  & 0.1799 & 0.2525 & 0.3604 \\
	SAR-DRN  & 25.25  & 28.39  & 29.77  & 0.5831 & 0.7143 & 0.8061 \\
	Ma2019 & 25.29 & 28.43 & 29.98 & 0.6237 & 0.7290 & 0.8059 \\
	PSD  & 25.47  & 28.45  & \textbf{30.04}  & 0.7023 & 0.7571 & 0.8200 \\
	PSDi  & \textbf{25.67}  & \textbf{28.67}  & 29.92  & \textbf{0.7196} & \textbf{0.7763} & \textbf{0.8284} \\
	\bottomrule
	\end{tabular}
  \label{synthetictable}
\end{table}
\vspace{0.2cm}

\label{sec:pagestyle}
\subsection{Experimental Setup}
In this work, 2.2 $\times$ ${\text{10}^\text{4}}$ one-look SAR image patches (cropped to 96 $\times$ 96 pixels) obtained from the Sentinel-1 are used to train the proposed networks.
In the training process of the adversarial learning (see Section~\ref{Generating}), to keep the noise level in the original one-look SAR images $\textbf{Y}$ consistent with that in the generated speckled images $\hat {\textbf{Y}}$, the number of looks $L$ for each generated speckled image is set to be 1.
In the training process of the N2N despeckling (see Section~\ref{Iterative}), to make the network available to different speckle levels, the number of looks $L$ for $( {\hat {\textbf{Y}}_1,\hat {\textbf{Y}}_2} )$ and $( {\tilde {\textbf{Y}}_1,\tilde {\textbf{Y}}_2} )$ is randomly set to be 1, 2, 4, 8 and 16. The predefined weight $\alpha$ is set to be 0.1. The networks ${g_{{\theta _{1}}}}$, ${g_{{\theta _{2}}}}$, ${g_{{\theta _{d}}}}$, and ${g_{{\theta _{di}}}}$ are trained using the Adam optimizer with $\beta_{1}=$0.9, $\beta_{2}=$0.999, and $\epsilon=$ ${\text{10}^{-\text{8}}}$, where the learning rates are all initialized as ${\text{10}^{-\text{4}}}$ and reduced by half with each eight epochs. The network $f_\omega$ is trained using the RMSProp optimizer with an initial learning rate of 5 $\times$ $\text{10}^{-\text{5}}$. Following~\cite{arjovsky2017wasserstein}, the clipping value is set as 0.02, and the critic iteration is 5.
The training process is set to have 16 epochs with mini-batch size 16 in both adversarial learning and N2N despeckling. PyTorch is employed to train the proposed methods with an Intel Xeon E5 CPU and an Nvidia 2080Ti GPU. The total training costs around 9 h 30 min.

\begin{figure*}[!t]
  \centering
  \subfloat[]{\includegraphics[width=0.73in]{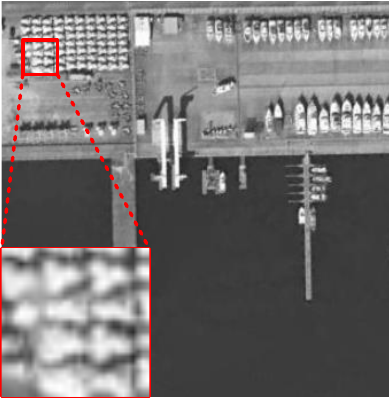}}\!
  \subfloat[]{\includegraphics[width=0.73in]{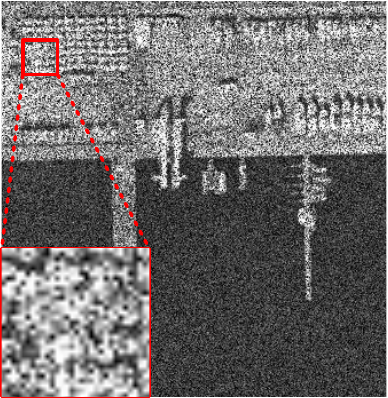}}\!
  \subfloat[]{\includegraphics[width=0.73in]{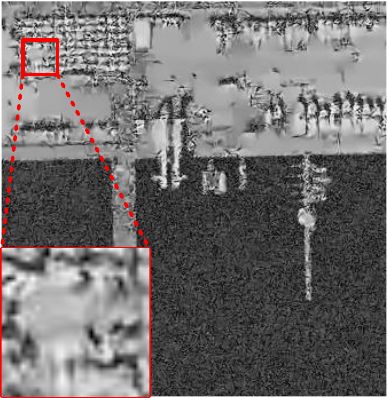}}\!
  \subfloat[]{\includegraphics[width=0.73in]{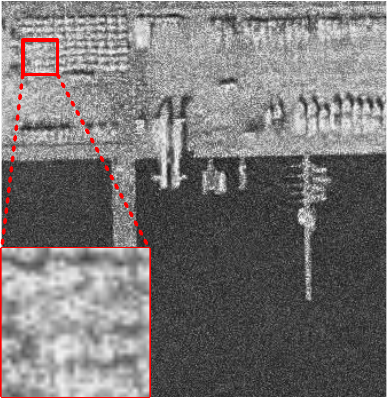}}\!
  \subfloat[]{\includegraphics[width=0.73in]{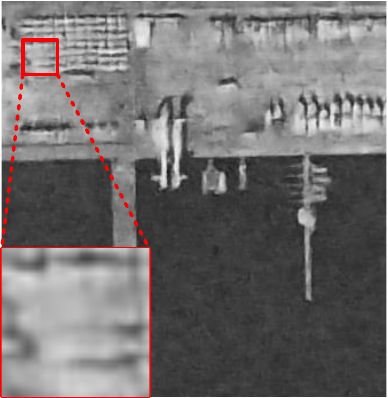}}\! 
  \subfloat[]{\includegraphics[width=0.73in]{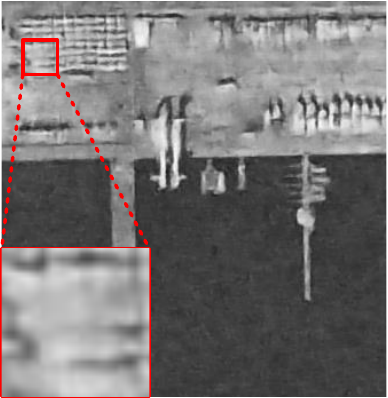}}\! 
  \subfloat[]{\includegraphics[width=0.73in]{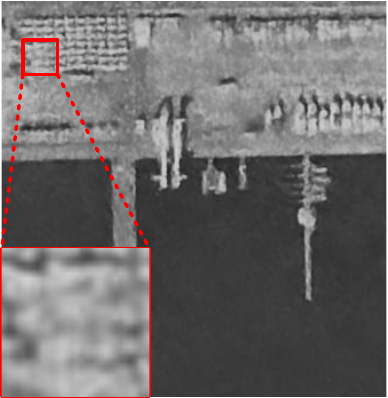}}\!
  \subfloat[]{\includegraphics[width=0.73in]{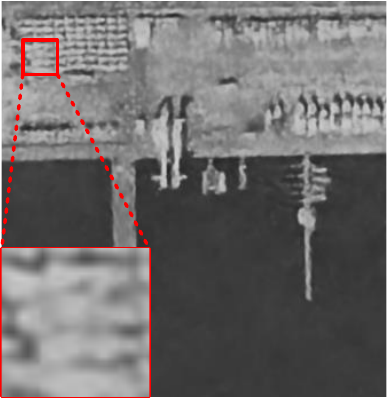}}\!
  \caption{Illustration of despeckling results on synthetic speckled image corrupted by four-look speckle. (a) Clean reference. (b) Speckled. (c) Chen2014. (d) SAR-BM3D. (e) SAR-DRN. (f) Ma2019. (g) PSD. (h) PSDi.}
  \label{AID}
  \vspace{-0.1cm}
\end{figure*}

\begin{figure*}[!t]
  \centering
  \includegraphics[width=0.77in]{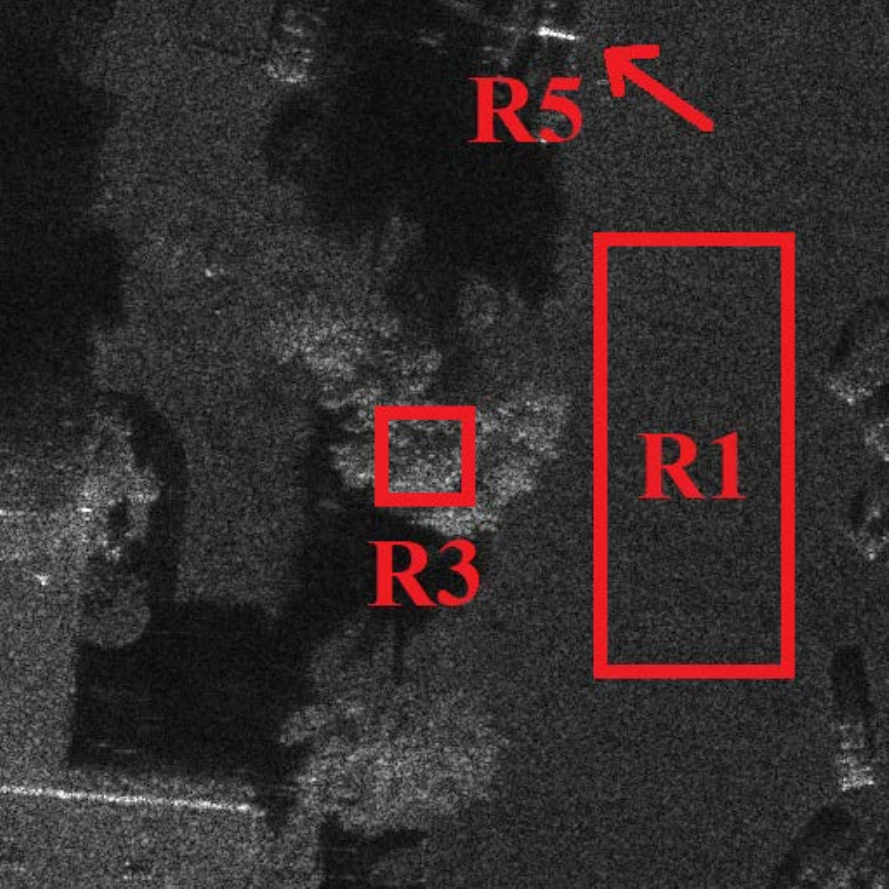}\!
  \includegraphics[width=0.77in]{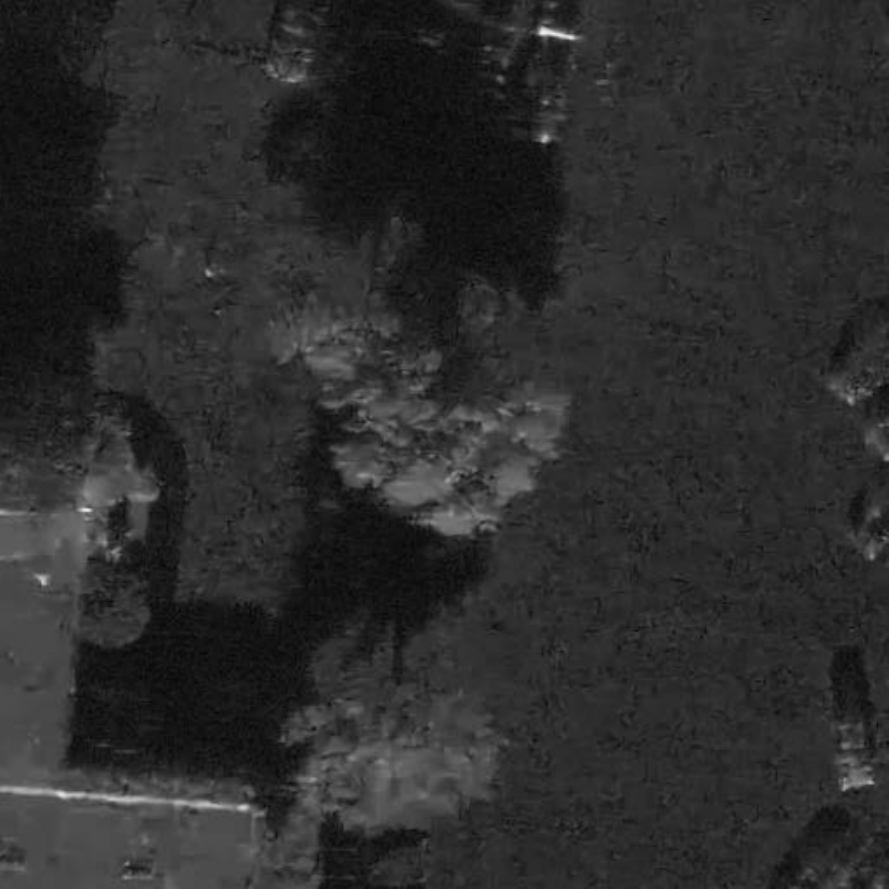}\!
  \includegraphics[width=0.77in]{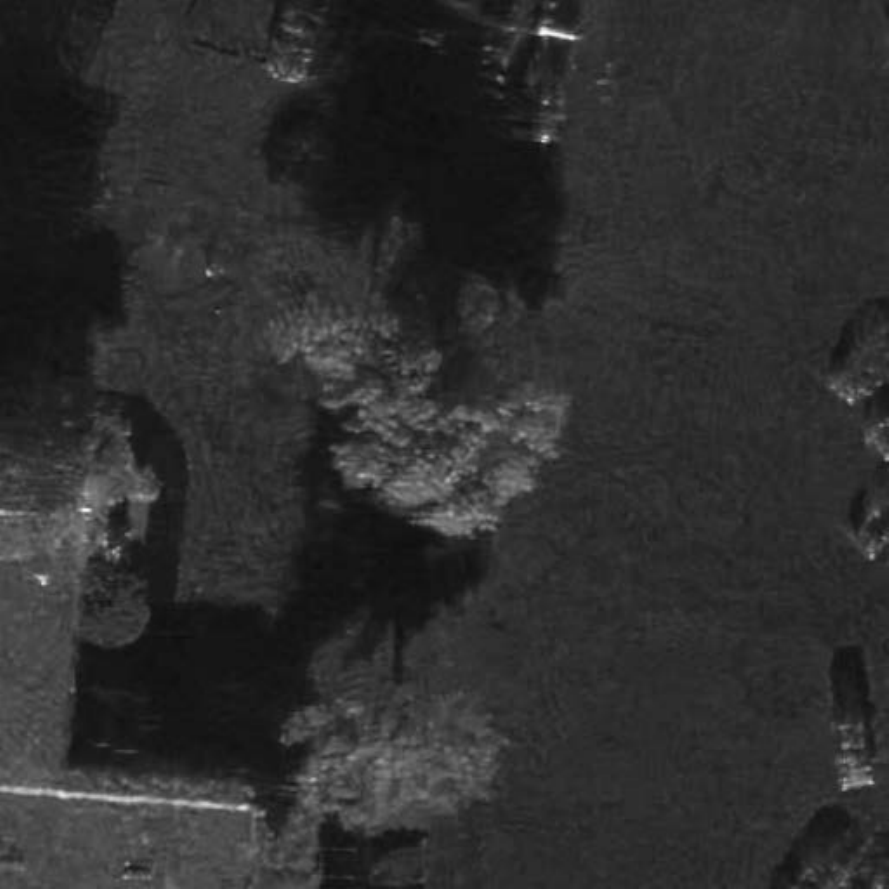}\!
  \includegraphics[width=0.77in]{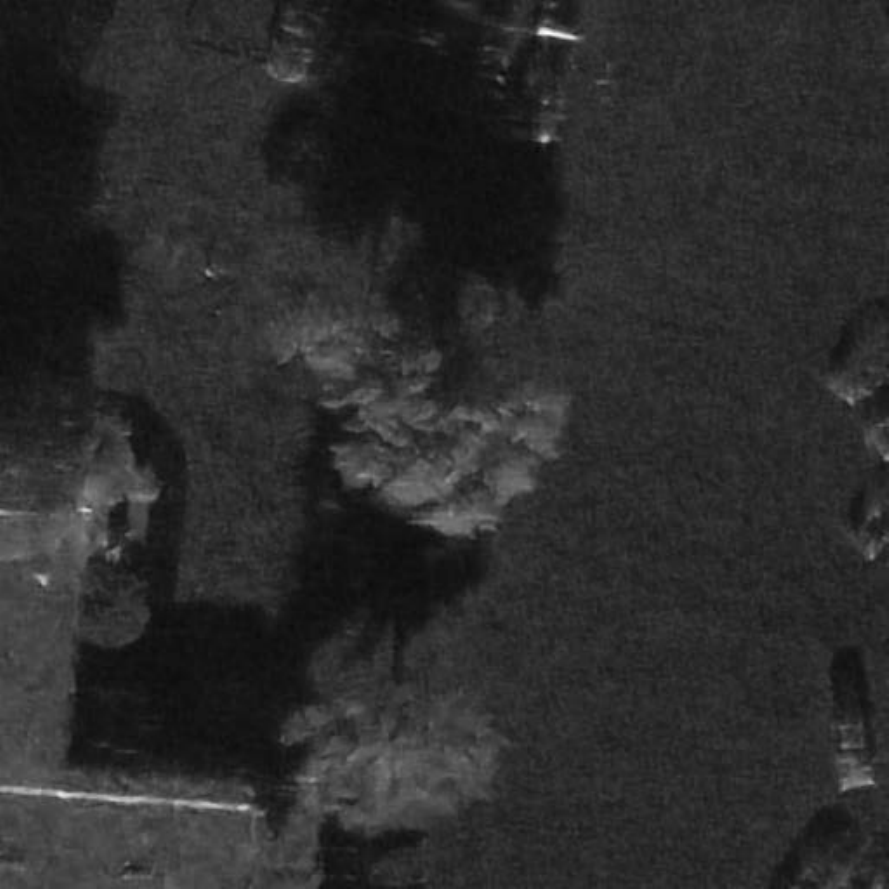}\!
  \includegraphics[width=0.77in]{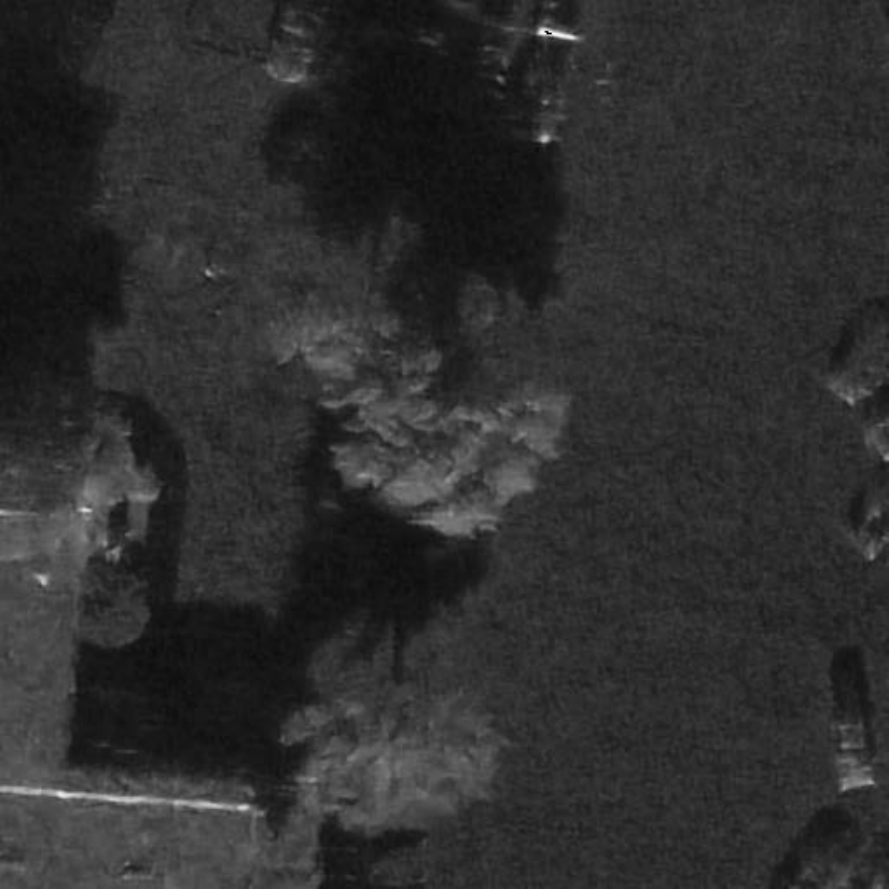}\!
  \includegraphics[width=0.77in]{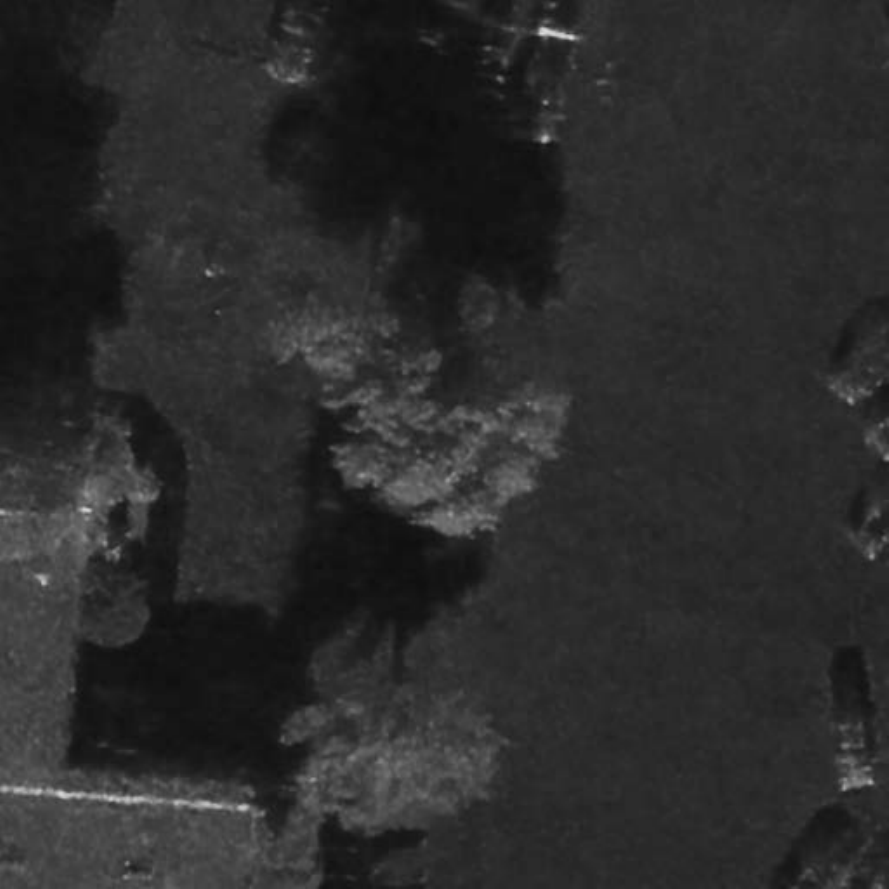}\!
  \includegraphics[width=0.77in]{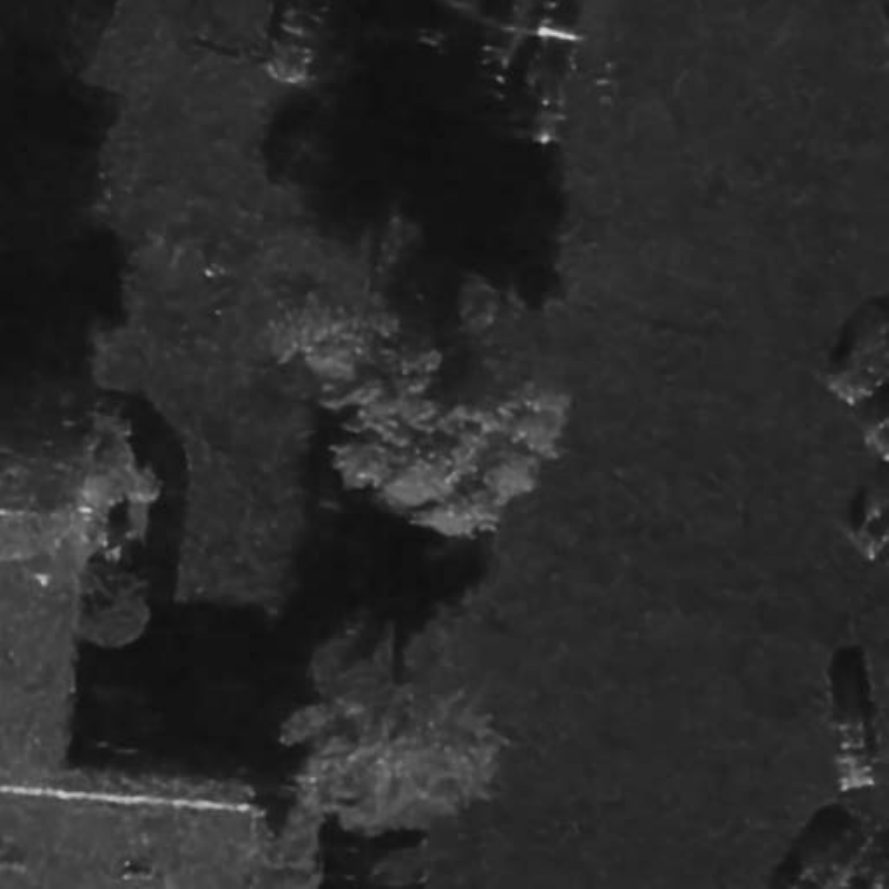}\!

  \vspace{-0.2cm}

  \subfloat[]{\includegraphics[width=0.77in]{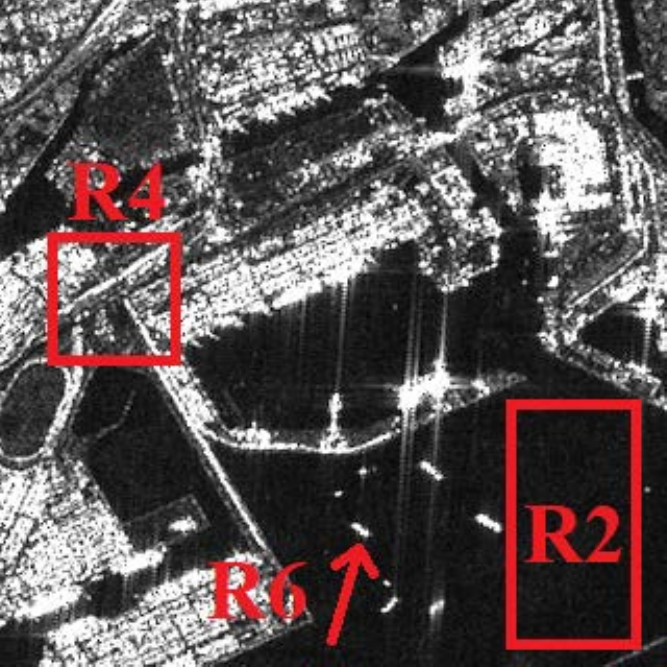}}\!
  \subfloat[]{\includegraphics[width=0.77in]{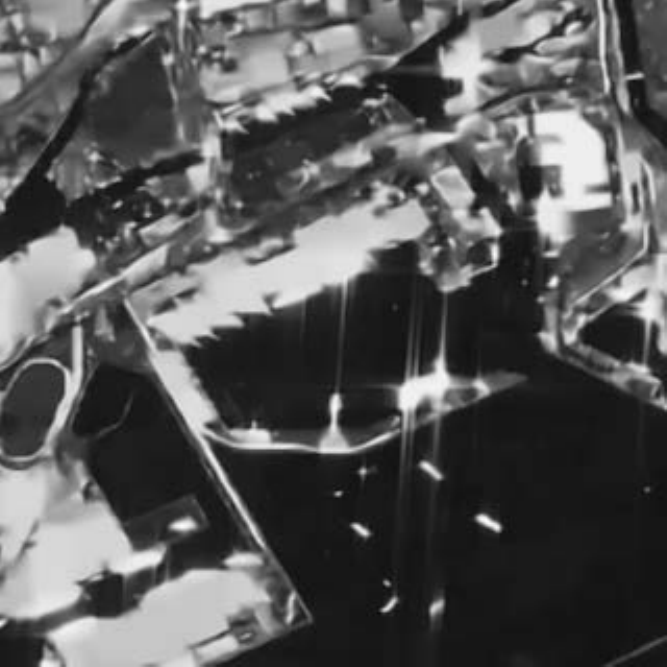}}\!
  \subfloat[]{\includegraphics[width=0.77in]{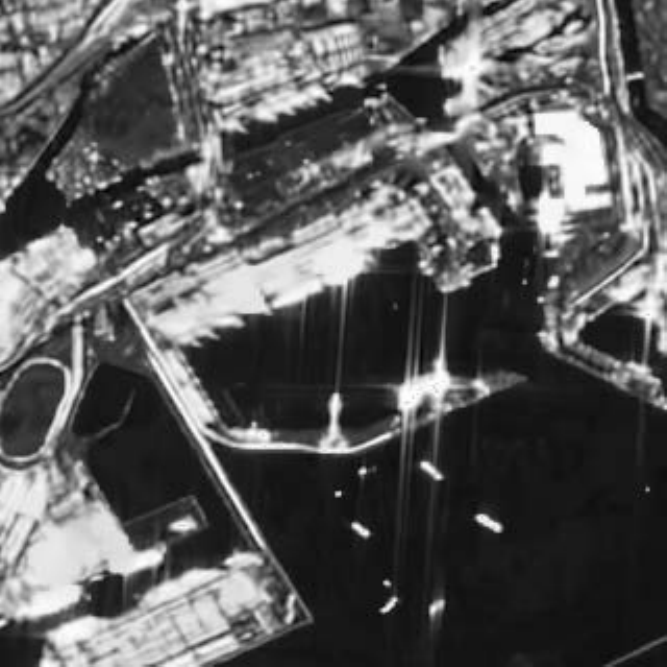}}\!
  \subfloat[]{\includegraphics[width=0.77in]{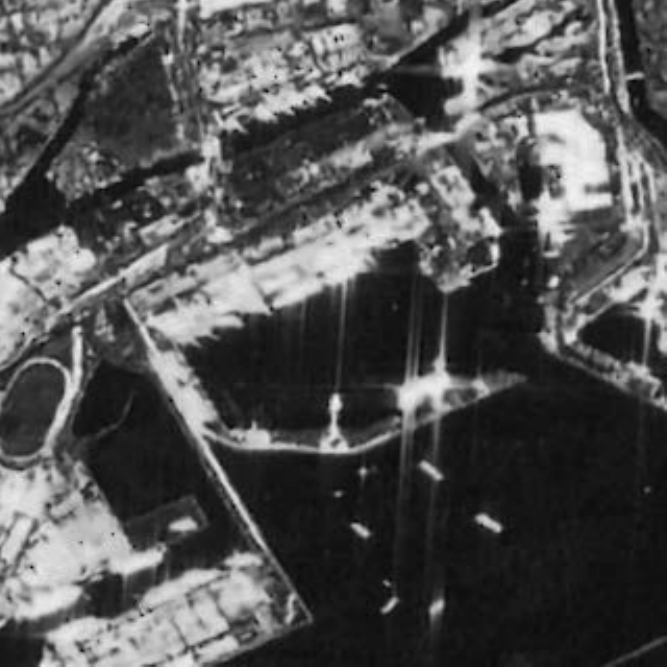}}\! 
  \subfloat[]{\includegraphics[width=0.77in]{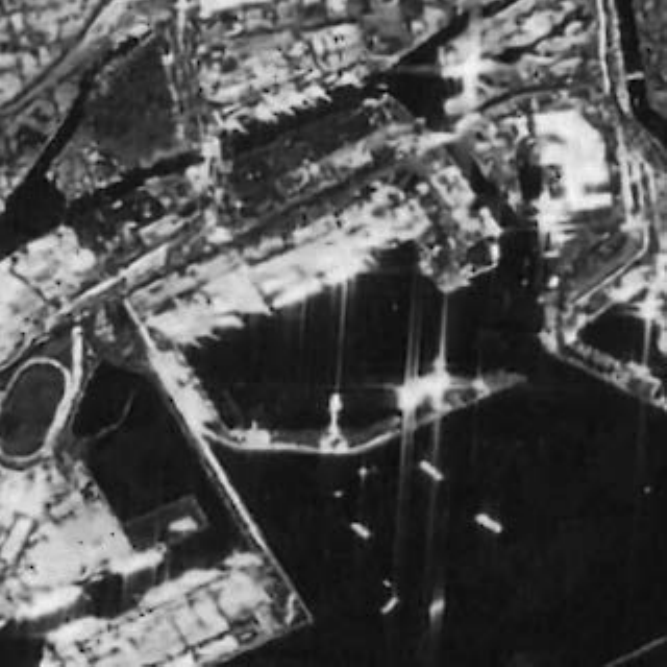}}\! 
  \subfloat[]{\includegraphics[width=0.77in]{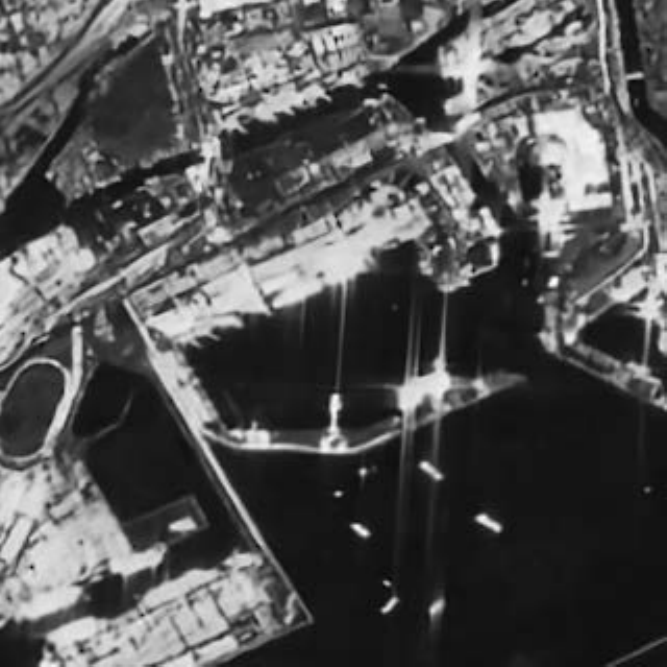}}\!
  \subfloat[]{\includegraphics[width=0.77in]{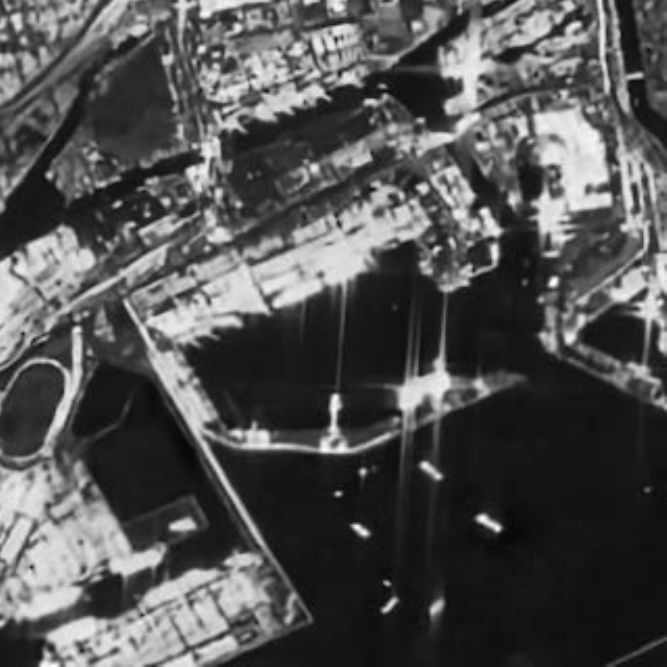}}\!
  \caption{(Top) Illustration of despeckling results on a three-look miniSAR image and (Bottom) a one-look Sentinel-1 SAR image. As identified by the red boxes and arrows, two homogeneous regions (R1 and R2) are employed to calculate ENL and MoR, two edge-feature regions (R3 and R4) are employed to calculate EPD-ROA, and two-point target regions (R5 and R6) are employed to calculate TCR. (a) Original image. (b) Chen2014. (c) SAR-BM3D. (d) SAR-DRN. (e) Ma2019. (f) PSD. (g) PSDi.}
  \label{Sandia}
  \vspace{-0.3cm}
\end{figure*}

To verify despeckling effectiveness, the proposed PSD (${g_{{{{\theta }}_d}}}$) and PSDi (${g_{{{{\theta }}_{di}}}}$) are compared with several state-of-the-art methods, i.e., variational-based Chen2014~\cite{chen2014higher}, nonlocal-based SAR-BM3D~\cite{parrilli_nonlocal_2012}, and SAR despeckling method with dilated residual network (SAR-DRN)~\cite{zhang_learning_2018} and Ma2019~\cite{ma2019cnn}. 
Here, though Chen2014 and SAR-BM3D do not need any additional training, they have to know the number of looks $L$, 
whereas SAR-DRN and Ma2019 need clean ground truth images for training, but do not need to know $L$ for testing. Hence, to train SAR-DRN and Ma2019, $2.2 \times 10^{4}$ speckle-free optical images with single channel from ImageNet~\cite{ILSVRC15} are used as the ground-truth images, and the input speckled images can be obtained by adding speckle to the ground truth images according to  \eqref{eq:1} and \eqref{eq:1-1}. To make a fair comparison, the number of looks $L$ is also randomly set to be 1, 2, 4, 8, and 16. Other parameters in SAR-DRN and Ma2019 are set as suggested in \cite{zhang_learning_2018} and \cite{ma2019cnn}, respectively.
\vspace{-0.3cm}
\subsection{Performance Comparison on Synthetic Speckled Data}
In this experiment, 100 optical remote sensing images (selected from Aerial Image data set (AID)~\cite{xia2017aid}) are used for analyzing. To verify the despeckling effectiveness with the known noise level, the number of looks $L$ of these images is set as 1, 4, and 16. Since the clean ground-truth images are available, the indices of peak signal to noise ratio (PSNR) and structural similarity index (SSIM)~\cite{wang2004image} are used for evaluation, and the comparison results are given in Table~\ref{synthetictable}. 
Chen2014 is performed following~\cite{chen2014higher}, as despeckling for 16-look images is not provided in \cite{chen2014higher}; to make a fair comparison, we only compare the despeckling results on one- and four-look images with Chen2014.
As can be seen from Table~\ref{synthetictable}, our proposed methods outperform other methods in all noise levels. Moreover, the PSDi method can further improve the despeckling performance compared with the PSD method.
Here, the highest SSIM values obtained by the proposed methods indicate the best structural features preservation, which can be beneficial for the applications, such as change detection with SAR images ~\cite{wang2019can}. As the visual comparison shown in Fig.~\ref{AID}, only our proposed methods can remove speckle effectively.

\begin{table}[tbp]
	\caption{Performance Comparison in Terms of ENL, EPD-ROA, TCR, and MoR on Real SAR Data}
	\centering
	\scriptsize
	\begin{tabular}{m{1.3cm}<{\centering}m{0.35cm}<{\centering}m{0.35cm}<{\centering}m{0.45cm}<{\centering}m{0.45cm}<{\centering}m{0.45cm}<{\centering}m{0.45cm}<{\centering}m{0.45cm}<{\centering}m{0.45cm}<{\centering}}
		\toprule
		\multirow{2}{*}{Data} & \multicolumn{2}{c}{ENL} & \multicolumn{2}{c}{EPD-ROA} & \multicolumn{2}{c}{TCR} & \multicolumn{2}{c}{MoR}  \\ 
		\cmidrule(r){2-3} \cmidrule(r){4-5} \cmidrule(r){6-7} \cmidrule(r){8-9}
		& R1 & R2 & R3 & R4 & R5 & R6 & R1 & R2 \\ \midrule
		Original & 12.40 & 4.61 & - & - & - & - & - & - \\ 
		Chen2014 & 66.78 & 104.26 & 0.8914 & 0.8404 & 1.3054 & 0.7723 & 0.9793 & 0.9736 \\ 
		SAR-BM3D & 138.72 & 87.17 & 0.8917 & 0.8546 & 1.1303 & 0.7707 & 0.9590 & 0.9417 \\ 
		SAR-DRN & 79.34 & 52.18 & 0.8928 & 0.8526 & 0.3370 & 0.4880 & 1.0305 & 1.0569 \\ 
		Ma2019 & 81.28 & 69.88 & 0.8931 & 0.8524 & 0.3332 & 0.6070 & 0.9484 & 0.9529 \\
		PSD & 203.45 & 134.92 & 0.8980 & 0.8547 & 0.2356 & \textbf{0.3651 
		} & \textbf{1.0152} & \textbf{1.0109} \\
		PSDi & \textbf{259.71} & \textbf{161.98} & \textbf{0.9018} & \textbf{0.8564} & \textbf{0.2348} & 0.3945 & 1.0196 & 1.0122 \\
		\bottomrule
		\label{ENL}
	\end{tabular}
	\vspace{-0.3cm}
\end{table}

\subsection{Performance comparison on real SAR data}
We also conduct despeckling experiments on real SAR data, where a three-look Ku-band miniSAR~\footnote{Courtesy of Sandia National Laboratories, Radar ISR.} image and a one-look C-band Sentinel-1 image are employed for evaluation. These images are cropped to 512 $\times$ 512 pixels and 320 $\times$ 320 pixels, respectively.
The results are given in Fig.~\ref{Sandia}. Equivalent number of looks (ENL)~\cite{parrilli_nonlocal_2012}, edge-preservation degree based on the ratio of average (EPD-ROA)~\cite{feng2011sar}, target-to-clutter ratio (TCR)~\cite{ma2017review}, and mean of ratio (MoR)~\cite{parrilli_nonlocal_2012} are used as performance metrics to evaluate the degree of speckle reduction, edge preservation, point target preservation, and radiometric preservation, respectively~\cite{ma2017review}. The results are given in Table~\ref{ENL}.

{As can be seen from Table~\ref{ENL}, SAR-BM3D and our proposed methods show better performance on speckle suppression. However, SAR-BM3D cannot show comparable performance as the proposed methods in preserving point target and radiometric.
SAR-DRN and Ma2019 are trained based on supervised learning by using optical images and corresponding synthetic speckled images, which are quite different from real SAR images. This leads to the domain gap problem, which makes networks performed well on the data (i.e., synthetic images) in the same domain as the training data, but poorly performed on real data. This can be observed from the comparison results from Table~\ref{synthetictable} and Table~\ref{ENL}, i.e., SAR-DRN and Ma2019 cannot show high speckle suppression ability on real SAR images as that on synthetic speckled images. In contrast, it is worth noting that our proposed methods can learn despeckling by training the network on real SAR data. 
Hence, our proposed methods can achieve better despeckling performance on both synthetic speckled data and real SAR data. Meanwhile, our proposed methods show well features preservation ability.} 
Especially, the proposed methods can obtain well tradeoff between EPD-ROA and ENL, which can be beneficial to agricultural area classification~\cite{lukin2018despeckling} and wetland classification~\cite{mahdianpari2017effect}.

In addition, we report the CPU time consumption of each method with the same system configuration, and the results are listed in Table~\ref{Runtime}. Similar to other deep learning-based despeckling methods (i.e., SAR-DRN and Ma2019), once the networks are well trained, the despeckling process for a given SAR image is very quick. Our proposed methods have achieved superior despeckling performance with comparable time consumption, compared with SAR-DRN and Ma2019. 

\begin{table}[tbp]
	\caption{{Time Consumption Comparison on Images With 256$\times$256 Pixels}}
	\centering
	\scriptsize
	\begin{tabular}{p{1cm}<{\centering}p{1cm}<{\centering}p{1cm}<{\centering}p{1cm}<{\centering}p{1cm}<{\centering}p{1cm}<{\centering}}
	\toprule
	Chen2014& \tabincell{c}{SAR-\\BM3D}  & \tabincell{c}{SAR-\\DRN}  &Ma2019& PSD & PSDi\\
	\midrule
	6.81s&18.42s&1.40s&\textbf{0.87s}&2.20s&2.21s\\
	\bottomrule
	\end{tabular}
	\label{Runtime}
	\vspace{-0.3cm}
\end{table}

\section{Conclusion}
We presented a practical deep learning-based methods for SAR image despeckling, which can achieve despeckling with only single speckled SAR images. Experiments conducted on both synthetic speckled data and real SAR data demonstrated the superiority of our methods compared with state-of-the-art methods. In our future study, we will explore the impact of speckle reduction and SAR feature preservation on applications using SAR images.

\ifCLASSOPTIONcaptionsoff
  \newpage
\fi



\bibliographystyle{IEEEtran}
\bibliography{PSDbibfile}

\begin{thebibliography}{10}
\providecommand{\url}[1]{#1}
\csname url@samestyle\endcsname
\providecommand{\newblock}{\relax}
\providecommand{\bibinfo}[2]{#2}
\providecommand{\BIBentrySTDinterwordspacing}{\spaceskip=0pt\relax}
\providecommand{\BIBentryALTinterwordstretchfactor}{4}
\providecommand{\BIBentryALTinterwordspacing}{\spaceskip=\fontdimen2\font plus
\BIBentryALTinterwordstretchfactor\fontdimen3\font minus
  \fontdimen4\font\relax}
\providecommand{\BIBforeignlanguage}[2]{{%
\expandafter\ifx\csname l@#1\endcsname\relax
\typeout{** WARNING: IEEEtran.bst: No hyphenation pattern has been}%
\typeout{** loaded for the language `#1'. Using the pattern for}%
\typeout{** the default language instead.}%
\else
\language=\csname l@#1\endcsname
\fi
#2}}
\providecommand{\BIBdecl}{\relax}
\BIBdecl

\bibitem{lukin2018despeckling}
V.~Lukin \emph{et~al.}, ``Despeckling of multitemporal sentinel {SAR} images
  and its impact on agricultural area classification,'' in \emph{Recent
  Advances and Applications in Remote Sensing}, London, U.K.: IntechOpen, 2018,
  pp. 3712--3719.

\bibitem{mahdianpari2017effect}
M.~Mahdianpari, B.~Salehi, and F.~Mohammadimanesh, ``The effect of {PolSAR}
  image despeckling on wetland classification: Introducing a new adaptive
  method,'' \emph{Can. J. Remote Sens.}, vol.~43, no.~5, pp. 485--503, Sep.
  2017.

\bibitem{wang2019can}
R.~Wang, J.-W. Chen, L.~Jiao, and M.~Wang, ``How can despeckling and structural
  features benefit to change detection on bitemporal {SAR} images?''
  \emph{Remote Sens.}, vol.~11, no.~4, p. 421, Feb. 2019.

\bibitem{lee_digital_1980}
J.-S. Lee, ``Digital image enhancement and noise filtering by use of local
  statistics,'' \emph{{IEEE} Trans. Pattern Anal. Mach. Intell.}, vol. PAMI-2,
  no.~2, pp. 165--168, Mar. 1980.

\bibitem{kuan1985adaptive}
D.~T. Kuan, A.~A. Sawchuk, T.~C. Strand, and P.~Chavel, ``Adaptive noise
  smoothing filter for images with signal-dependent noise,'' \emph{{IEEE}
  Trans. Pattern Anal. Mach. Intell.}, vol. PAMI-7, no.~2, pp. 165--177, Mar.
  1985.

\bibitem{chang2000adaptive}
S.~G. Chang, B.~Yu, and M.~Vetterli, ``Adaptive wavelet thresholding for image
  denoising and compression,'' \emph{IEEE Trans. Image Process.}, vol.~9,
  no.~9, pp. 1532--1546, Sep. 2000.

\bibitem{li2012bayesian}
H.-C. Li, W.~Hong, Y.-R. Wu, and P.-Z. Fan, ``Bayesian wavelet shrink-age with
  heterogeneity-adaptive threshold for {SAR} image despeckling based on
  generalized gamma distribution,'' \emph{IEEE Trans. Geosci. Remote Sens.},
  vol.~51, no.~4, pp. 2388--2402, Apr. 2013.

\bibitem{aubert2008variational}
G.~Aubert and J.-F. Aujol, ``A variational approach to removing mul-tiplicative
  noise,'' \emph{SIAM J. Appl. Math.}, vol.~68, no.~4, pp. 925--946, Jan. 2008.

\bibitem{shi2008nonlinear}
J.~Shi and S.~Osher, ``A nonlinear inverse scale space method for a convex
  multiplicative noise model,'' \emph{SIAM J. Appl. Math.}, vol.~1, no.~3, pp.
  294--321, Jan. 2008.

\bibitem{chen2014higher}
Y.~Chen, W.~Feng, R.~Ranftl, H.~Qiao, , and T.~Pock, ``A higher-order {MRF}
  based variational model for multiplicative noise reduction,'' \emph{IEEE
  Signal Process. Lett.}, vol.~21, no.~11, pp. 1370--1374, Nov. 2014.

\bibitem{parrilli_nonlocal_2012}
S.~Parrilli, M.~Poderico, C.~V. Angelino, , and L.~Verdoliva, ``A nonlocal
  {SAR} image denoising algorithm based on {LLMMSE} wavelet shrink-age,''
  \emph{{IEEE} Trans. Geosci. Remote Sens.}, vol.~50, no.~2, pp. 606--616, Feb.
  2012.

\bibitem{argenti2013tutorial}
F.~Argenti, A.~Lapini, T.~Bianchi, and L.~Alparone, ``A tutorial on speckle
  reduction in synthetic aperture radar images,'' \emph{IEEE Geosci. Remote
  Sens. Mag.}, vol.~1, no.~3, pp. 6--35, Sep. 2013.

\bibitem{chierchia_sar_2017}
G.~Chierchia, D.~Cozzolino, G.~Poggi, and L.~Verdoliva, ``{SAR} image
  despeckling through convolutional neural networks,'' in \emph{{Proc. {IEEE}
  Int. Geosci. Remote Sens. Symp. (IGARSS)}}, Jul. 2017, pp. 5438--5441.

\bibitem{wang_sar_2017}
P.~Wang, H.~Zhang, and V.~M. Patel, ``{SAR} image despeckling using a
  convolutional neural network,'' \emph{{IEEE} Signal Process. Lett.}, vol.~24,
  no.~12, pp. 1763--1767, Dec. 2017.

\bibitem{zhang_learning_2018}
Q.~Zhang, Q.~Yuan, J.~Li, Z.~Yang, and X.~Ma, ``Learning a dilated residual
  network for {SAR} image despeckling,'' \emph{Remote Sens.}, vol.~10, no.~2,
  p. 196, Jan. 2018.

\bibitem{ma2019cnn}
D.~Ma, X.~Zhang, X.~Tang, J.~Ming, and J.~Shi, ``A {CNN}-based method for {SAR}
  image despeckling,'' in \emph{{Proc. {IEEE} Int. Geosci. Remote Sens. Symp.
  (IGARSS)}}, Jul. 2019, pp. 4272--4275.

\bibitem{di2013benchmarking}
G.~D. Martino, M.~Poderico, G.~Poggi, D.~Riccio, and L.~Verdoliva,
  ``Benchmarking framework for {SAR} despeckling,'' \emph{IEEE Trans. Geosci.
  Remote Sens.}, vol.~52, no.~3, pp. 1596--1615, Mar. 2014.

\bibitem{Lehtinen2018Noise2NoiseLI}
J.~Lehtinen \emph{et~al.}, ``{Noise2Noise}: {Learning} image restoration
  without clean data,'' in \emph{Proc. Int. Conf. Mach. Learn.}, 2018, pp.
  2965--2974.

\bibitem{zhou2018unetjia}
Z.~Zhou, M.~Siddiquee, N.~Tajbakhsh, and J.~Liang, ``{UNet++:} {A} nested
  {U-Net} architecture for medical image segmentation,'' in \emph{Proc. Int.
  Workshop Deep Learn. Med. Image Anal. Int. Workshop Multimodal Learn. Clin.
  Decis. Support}, 2018, pp. 3--11.

\bibitem{cha2019gan2gan}
\BIBentryALTinterwordspacing
S.~Cha, T.~Park, and T.~Moon, ``{GAN2GAN}: {Generative} noise learning for
  blind image denoising with single noisy images,'' \emph{arXiv:1905.10488},
  2019. [Online]. Available: \url{http://arxiv.org/abs/1905.10488}
\BIBentrySTDinterwordspacing

\bibitem{arjovsky2017wasserstein}
M.~Arjovsky, S.~Chintala, and L.~Bottou, ``Wasserstein generative adversarial
  networks,'' in \emph{Proc. 34th Int. Conf. Mach. Learn.}, 2017, pp. 214--223.

\bibitem{zhang2017beyond}
K.~Zhang, W.~Zuo, Y.~Chen, D.~Meng, and L.~Zhang, ``Beyond a gaussian denoiser:
  {Residual} learning of deep {CNN} for image denoising,'' \emph{IEEE Trans.
  Image Process.}, vol.~26, no.~7, pp. 3142--3155, Jul. 2017.

\bibitem{zhu2017unpaired}
J.-Y. Zhu, T.~Park, P.~Isola, and A.~A. Efros, ``Unpaired image-to-image
  translation using cycle-consistent adversarial networks,'' in \emph{Proc.
  {IEEE} Int. Conf. Comput. Vis. (ICCV)}, Oct. 2017, pp. 2223--2232.

\bibitem{lim_enhanced_2017}
B.~Lim, S.~Son, H.~Kim, S.~Nah, , and K.~M. Lee, ``Enhanced deep residual
  networks for single image super-resolution,'' in \emph{Proc. {IEEE} Int.
  Conf. Comput. Vis. Pattern Recognit. Workshop (CVPRW)}, Jul. 2017, pp.
  1132--1140.

\bibitem{ILSVRC15}
O.~Russakovsky \emph{et~al.}, ``{ImageNet} large scale visual recognition
  chal-lenge,'' \emph{Int. J. Comput. Vis.}, vol. 115, no.~3, pp. 211--252,
  Dec. 2015.

\bibitem{xia2017aid}
G.-S. Xia \emph{et~al.}, ``{AID}: {A} benchmark data set for performance
  evaluation of aerial scene classification,'' \emph{{IEEE} Trans. Geosci.
  Remote Sens.}, vol.~55, no.~7, pp. 3965--3981, Jul. 2017.

\bibitem{wang2004image}
Z.~Wang, A.~C. Bovik, H.~R. Sheikh, and E.~P. Simoncelli, ``Image quality
  assessment: From error visibility to structural similarity,'' \emph{{IEEE}
  Trans. on Image Process.}, vol.~13, no.~4, pp. 600--612, Apr. 2004.

\bibitem{feng2011sar}
H.~Feng, B.~Hou, and M.~Gong, ``{SAR} image despeckling based on local
  homogeneous-region segmentation by using pixel-relativity measure-ment,''
  \emph{IEEE Trans. Geosci. Remote Sens.}, vol.~7, no.~49, pp. 2724--2737, Jul.
  2011.

\bibitem{ma2017review}
X.~Ma, P.~Wu, Y.~Wu, and H.~Shen, ``A review on recent developments in fully
  polarimetric {SAR} image despeckling,'' \emph{IEEE J. Sel. Top. Appl. Earth
  Obs. Remote Sens.}, vol.~11, no.~3, pp. 743--758, Mar. 2018.

\end{thebibliography}
\end{document}